\documentclass[aps,prb,showpacs,preprint,floatfix]{revtex4-1}

\usepackage{graphicx}
\usepackage{dcolumn}
\usepackage{amsmath}
\usepackage{rotating}
\usepackage[latin1]{inputenc}

\begin{document}

\title{Orthorhombic to tetragonal transition of SrRuO$_3$ layers in Pr$_{0.7}$Ca$_{0.3}$MnO$_3$/SrRuO$_3$ superlattices}
\author{M. Ziese}
\email{ziese@physik.uni-leipzig.de}
\affiliation{Division of Superconductivity and Magnetism, University of Leipzig, D-04103 Leipzig, Germany}
\author{I. Vrejoiu}
\email{vrejoiu@mpi-halle.de}
\affiliation{Max Planck Institute of Microstructure Physics, D-06120 Halle, Germany}
\author{E. Pippel}
\affiliation{Max Planck Institute of Microstructure Physics, D-06120 Halle, Germany}
\author{E. Nikulina}
\affiliation{Max Planck Institute of Microstructure Physics, D-06120 Halle, Germany}
\author{D. Hesse}
\affiliation{Max Planck Institute of Microstructure Physics, D-06120 Halle, Germany}

\date{\today}

\begin{abstract}
High-quality Pr$_{0.7}$Ca$_{0.3}$MnO$_3$/SrRuO$_3$ superlattices with ultrathin layers were fabricated by pulsed laser deposition
on SrTiO$_3$ substrates. The superlattices were studied by atomically resolved scanning transmission electron
microscopy, high-resolution transmission electron microscopy, resistivity and magnetoresistance measurements.
The superlattices grew coherently without growth defects. Viewed along the growth direction, SrRuO$_3$ and Pr$_{0.7}$Ca$_{0.3}$MnO$_3$
layers were terminated by RuO$_2$ and MnO$_2$, respectively, which imposes a unique structure to their interfaces.
Superlattices with a constant thickness of the SrRuO$_3$ layers, but varying thickness of the Pr$_{0.7}$Ca$_{0.3}$MnO$_3$
layers showed a change of crystalline symmetry of the SrRuO$_3$ layers. At a low Pr$_{0.7}$Ca$_{0.3}$MnO$_3$ layer thickness of 1.5~nm
transmission electron microscopy proved the SrRuO$_3$ layers to be orthorhombic, whereas these were non-orthorhombic
for a Pr$_{0.7}$Ca$_{0.3}$MnO$_3$ layer thickness of 4.0~nm. Angular magnetoresistance measurements showed orthorhombic
(with small monoclinic distortion) symmetry in the first case and tetragonal symmetry of the SrRuO$_3$ layers in the second case.
Mechanisms driving this orthorhombic to tetragonal transition are briefly discussed.
\end{abstract}
\maketitle
\clearpage

\section{Introduction}
Heterostructures and superlattices (SLs) of oxide perovskites open an exciting field of research, since it is possible by
present epitaxy techniques to grow samples with clearly defined interfaces allowing for the realization of new functionalities.
Some examples are the two-dimensional electron gas at the SrTiO$_3$-LaAlO$_3$ interface,\cite{reyren2007} electron tunnelling in multiferroic
systems,\cite{gajek2007} growth of extrinsic multiferroic superlattices,\cite{vrejoiu2008} as well as the observation
of a giant interlayer exchange coupling in La$_{0.7}$Sr$_{0.3}$MnO$_3$/SrRuO$_3$ superlattices.\cite{ziese2010a,ziese2010b}
The latter exchange coupling leads to positive exchange bias\cite{ke2004,ke2005,padhan2006} and is mediated by the direct Mn-O-Ru-bond.
\cite{lee2008b,ziese2010a} The exchange bias is very large, since the individual layer thickness in the SLs is very small. In
general, in systems with strong correlations between the electronic, magnetic and structural degrees of freedom one
would expect to find new phenomena in the limit of ultrathin layers, i.e.~in restricted geometries.

In this paper, another type of perovskite superlattice is studied, consisting out of ultrathin Pr$_{0.7}$Ca$_{0.3}$MnO$_3$ (PCMO) and SrRuO$_3$ (SRO)
layers. Bulk SRO is an itinerant ferromagnet with orthorhombic crystal structure (space group Pbnm, lattice parameters
$a = 0.55670$~nm, $b = 0.55304$~nm, $c = 0.78446$~nm) and a Curie temperature of about 160~K.\cite{cao1997,klein1996a} Bulk PCMO
has orthorhombic symmetry (Pbnm, $a = 0.5426$~nm, $b = 0.5478$~nm, $c = 0.7679$~nm); it has a complex magnetic behavior and phase diagram,
and for 30\% Ca doping several transitions occur upon cooling, with an insulating canted ferromagnetic or antiferromagnetic state
below $\simeq 110$~K \cite{jirak1985,yoshizawa1995}.
The aim of this work is to investigate the crystalline symmetry of the individual layers. This is a formidable task, since the layer
thickness is below 5~nm and since the orthorhombic distortions from the pseudocubic cell are at maximum $0.3$\% for SRO and $0.6$\% for PCMO.
This task was tackled by high-resolution transmission electron microscopy as well as angular-dependent magnetoresistance (MR)
measurements. Since PCMO is insulating, the MR measurements only probe the SRO layers. We have shown before that the crystalline
symmetry of orthorhombic SRO single layers could be accurately studied by angular MR measurements, revealing
a monoclinic distortion of the $a$- and $b$-axes\cite{ziese2010c} that was also observed in high-resolution X-ray diffractometry.\cite{gan1999}
\section{Experimental\label{experimental}}
PCMO/SRO SLs were fabricated by pulsed-laser deposition at a temperature of $650^\circ$C and in an oxygen partial pressure
of $0.14$~mbar. Vicinal SrTiO$_3$ $(100)$ single crystal substrates with a low miscut angle of about $0.1^\circ$ were used for the growth,
after being etched in buffered HF and annealed at $1000^\circ$C for 2~hours in air. This treatment assured substrate surfaces with
atomically flat terraces of a width between 100 and 500~nm separated by unit-cell high steps. The SLs consisted of fifteen PCMO/SRO
bilayers with various layer thicknesses, see Table~\ref{table1}.
\begin{table}
\caption{Samples studied in this work. For all samples the Curie temperature of the SRO layers
was $T_C = 143$~K and the N\'eel temperature of the PCMO layers was $T_N = 110$~K.}
\label{table1}
\begin{tabular}{lll}
\hline
Sample & $[$PCMO / SRO$]_{15}$ & \\
\hline
SL1 & [1.5~nm / 4.4~nm] & [4 u.c. / 10-11 u.c.]\\
SL2 & [3.0~nm / 4.0~nm] & [8 u.c. / 9-10 u.c.]\\
SL3 & [3.8~nm / 4.0~nm] & [10 u.c. / 10 u.c.]\\
\hline
\end{tabular}
\end{table}

High-angle annular dark-field scanning transmission electron microscopy (HAADF-STEM), electron energy loss spectroscopy
(EELS) and energy dispersive X-ray (EDX) mappings were done in a TITAN 80-300 FEI microscope
(300~keV energy of the primary electrons) with a spherical aberration corrected ($c_s = 0$)
probe forming system. For the related Scherzer conditions\cite{scherzer1949} used, i.e.~a focus of  $\Delta = c_s = 0$~nm, image aberrations
were minimum and all atomic columns were clearly resolved in the HAADF-STEM mode.
High-resolution transmission electron microscopy (HRTEM) investigations were performed in a Jeol 4010 (400~keV energy of the primary electrons),
and Fourier-filter-related image processing was performed by help of the Digital Micrograph program package (Gatan Inc.).
For magnetoresistance measurements the SLs were mounted on a rotatable stage with an angular resolution better than
$0.01^\circ$ and an angle slackness after reversal of $0.1-0.2^\circ$.
The measurements were performed in a He-flow cryostat equipped with an 8~T superconducting solenoid.
\section{Structural properties\label{tem}}
\begin{figure}[t]
\centerline{\includegraphics[width=0.45\textwidth]{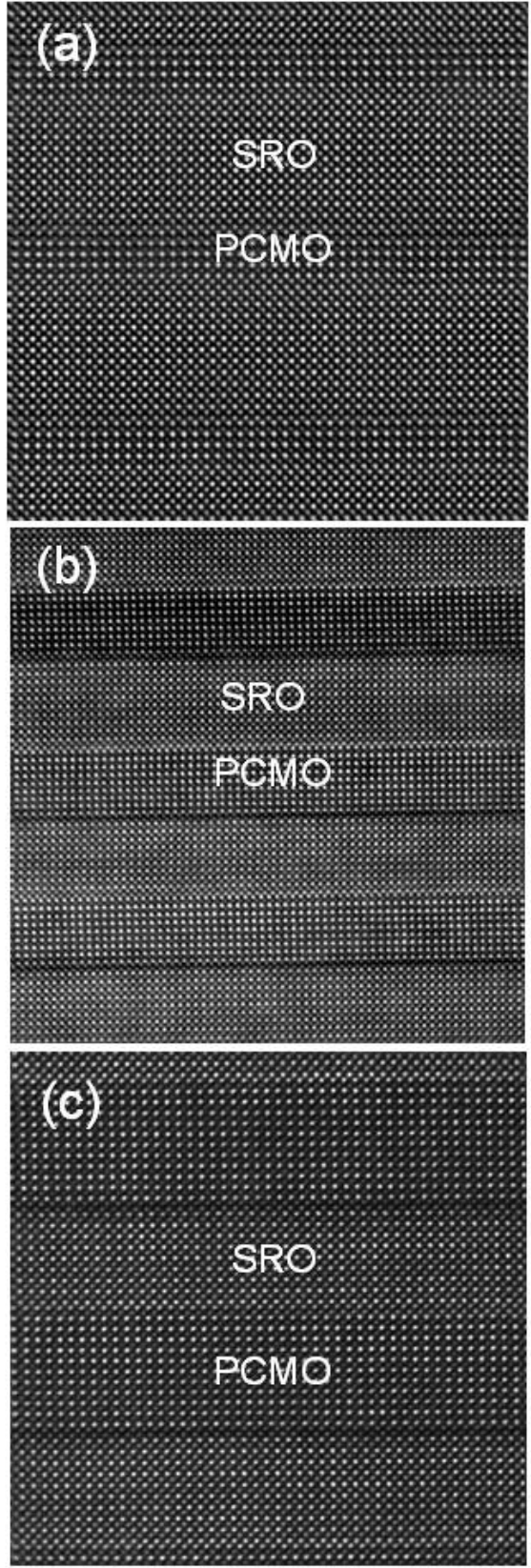}}
\caption{HAADF-STEM images of samples (a) SL1 (1.5~nm/4.4~nm), (b) SL2 (3.0~nm/4.0~nm) and (c) SL3 (3.8~nm/4.0~nm).}
\label{s1}
\end{figure}
\begin{figure}[t]
\centerline{\includegraphics[width=0.85\textwidth]{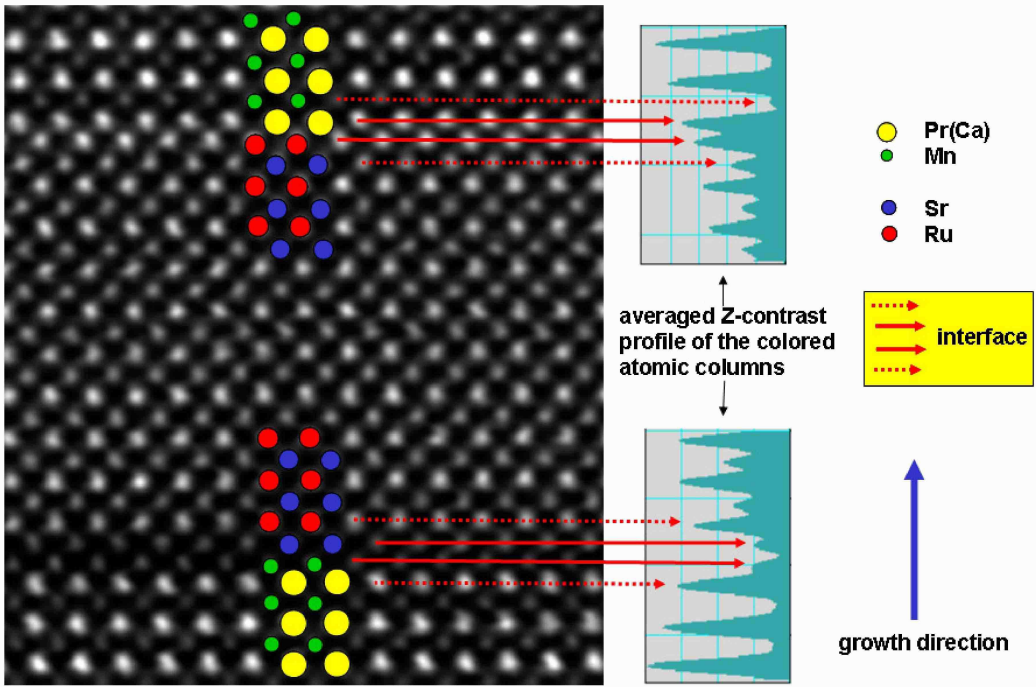}}
\caption{Z-STEM image of sample SL3 showing the interfacial structure. The intensity scans shown to the right
of the image allow for a unique determination of the cation species due to the monotonic
dependence of intensity on atomic number.}
\label{s2}
\end{figure}
\begin{figure}[t]
\centerline{\includegraphics[width=0.95\textwidth]{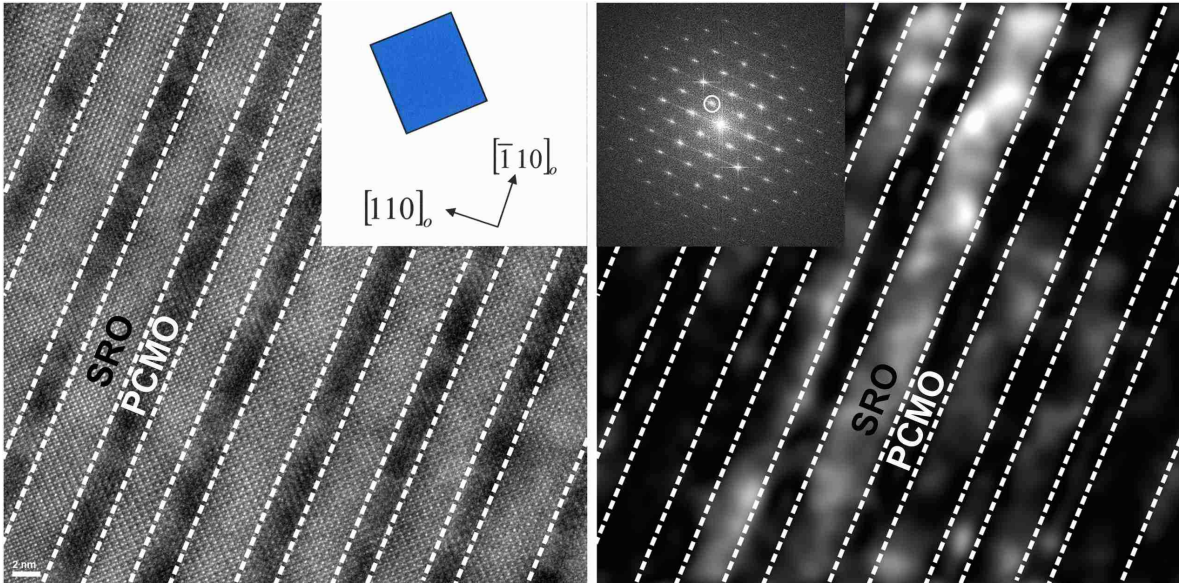}}
\caption{Sample SL1. Cross-sectional HRTEM image (left), fast-Fourier spectrum (right inset), reconstructed dark-field image in the light of
the $(010)_o$ reflection (right) and scheme of the oriented projection of the orthorhombic SRO unit cell with in-plane c-axis along the viewing
direction (left inset). Mind the scale bar (2~nm) in the bottom left corner.}
\label{s3}
\end{figure}
\begin{figure}[t]
\centerline{\includegraphics[width=0.95\textwidth]{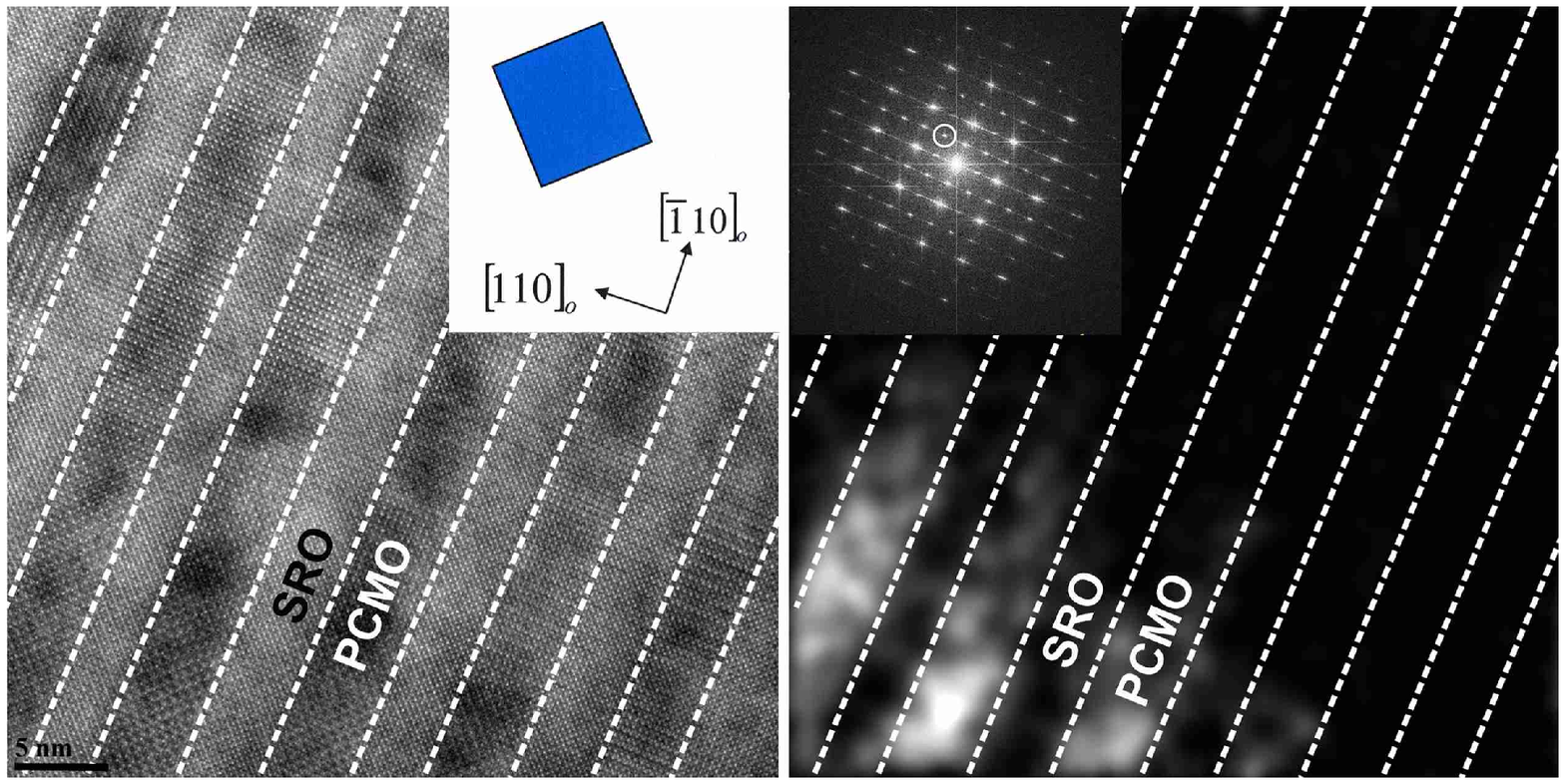}}
\caption{Sample SL3. Same as Fig.~\protect{\ref{s3}}.  Mind the scale bar (5~nm) in the bottom left corner.}
\label{s4}
\end{figure}
Figure~\ref{s1} shows HAADF-STEM micrographs of the three PCMO/SRO SLs, with respective layers thicknesses of 1.5~nm/4.4~nm (SL1, Fig.~\ref{s1}(a)),
3.0~nm/4.0~nm (SL2, Fig.~\ref{s1}(b)) and 3.8~nm/4.0~nm (SL3, Fig.~\ref{s1}(c)). The layers were grown entirely epitaxially, with coherent interfaces between
the PCMO and SRO layers. No misfit dislocations were found along the interfaces. Closer inspection of the HAADF-STEM micrographs revealed
an asymmetry of the interfaces: in the growth direction, the PCMO layers terminated most probably with MnO$_2$ planes, and the SRO layers
terminated most likely with RuO$_2$ planes, resulting in different interface contrasts, see Fig.~\ref{s2}.

Fast Fourier transforms (FFTs) of HRTEM and STEM micrographs showed orthorhombic reflections indicating that either PCMO or SRO, or both,
had orthorhombic structure in the SLs. Bulk PCMO and SRO have orthorhombic structures at room temperature, however, for epitaxial films, especially
coherent and ultrathin ones grown on dissimilar substrates, distortions from the orthorhombic bulk structure and formation of particular
configurations of crystallographic domains are expected to occur.\cite{fujimoto2007,gan1999} For example, epitaxial SRO films on DyScO$_3$$(110)$
were proven to have tetragonal structure.\cite{vailionis2008}

Dark-field reconstructed images in the light of certain reflections, obtained from cross- sectional HRTEM images of samples SL1 and SL3,
revealed a characteristic difference between these samples: whereas the SRO layers of sample SL1 were clearly orthorhombic, with the long
orthorhombic axis lying in the plane of the layers, the SRO layers in sample SL3 were either not orthorhombic or contain only very few
orthorhombic domains.

\begin{figure}[t]
\centerline{\includegraphics[width=0.95\textwidth]{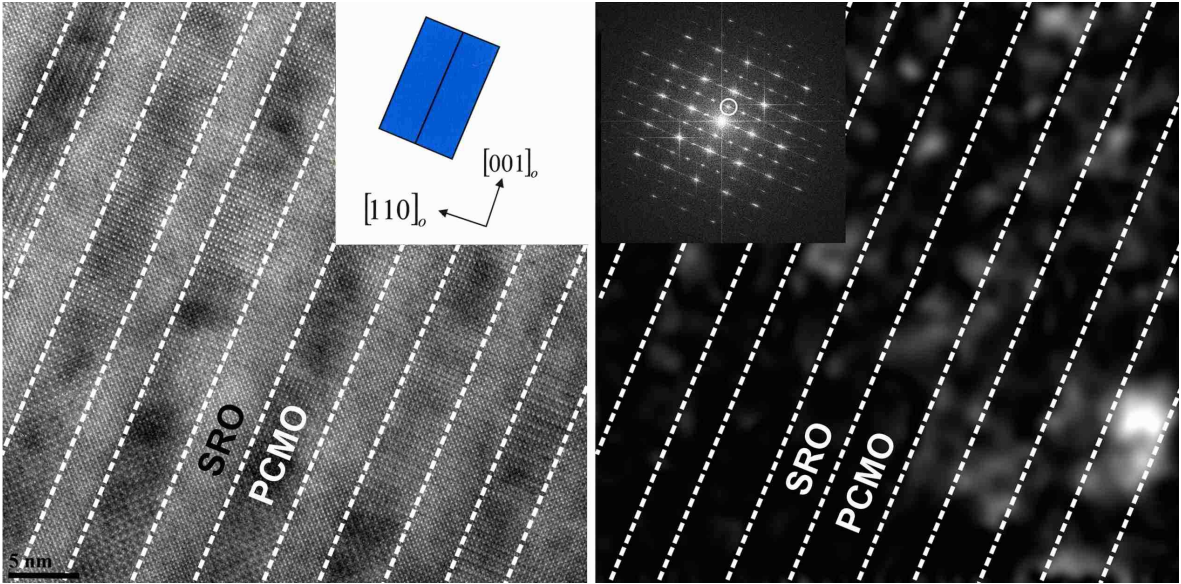}}
\caption{Sample SL3. Same as Fig.~\protect{\ref{s4}}, but with reconstruction in the light of the
$(001)_o$ reflection. {\em The in-plane c-axis is perpendicular to the viewing direction.}
Mind the scale bar (5~nm) in the bottom left corner.}
\label{s5}
\end{figure}

\begin{figure}[t]
\centerline{\includegraphics[width=0.95\textwidth]{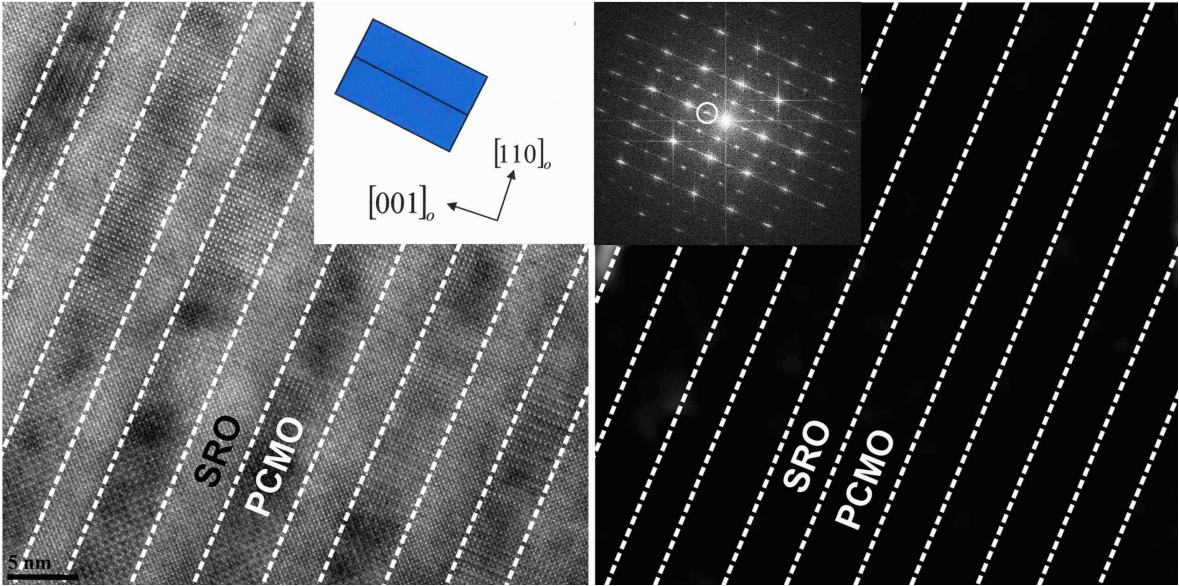}}
\caption{Sample SL3. Same as Fig.~\protect{\ref{s5}}, but {\em with the out-of-plane c-axis perpendicular to the viewing
direction.} Mind the scale bar (5~nm) in the bottom left corner.}
\label{s6}
\end{figure}

HRTEM images were taken from cross sections of samples SL1 and SL3. FFTs and reconstructed dark-field images in the light of certain
reflections were prepared. Note again that in the following the long orthorhombic axis
of the SRO unit cell is defined as the $c$-axis. In particular, the following reflections were used to characterize the superlattices with respect
to the presence of an orthorhombic phase in the SRO layers:

\begin{enumerate}
\item
the orthorhombic $(010)_o$ reflections corresponding to those parts of both SRO and PCMO lattices in which the orthorhombic $c$-axis was
potentially in-plane (i.e.~in the plane of the layers) along the viewing direction (Figs.~\ref{s3} and \ref{s4}, see insets);
\item
the orthorhombic $(001)_{o}$ reflections corresponding to those parts of both SRO and PCMO lattices in which the orthorhombic
$c$-axis was potentially in-plane but perpendicular to the viewing direction (Fig.~\ref{s5}, see inset), and
\item
the orthorhombic $(001)_{o}$ reflections corresponding to those parts of both SRO and PCMO lattices in which the orthorhombic
$c$-axis was potentially out-of-plane (perpendicular to the plane of the layers and perpendicular to the viewing direction)
(Fig.~\ref{s6}, see inset).
\end{enumerate}

\begin{figure}[t]
\centerline{\includegraphics[width=0.55\textwidth]{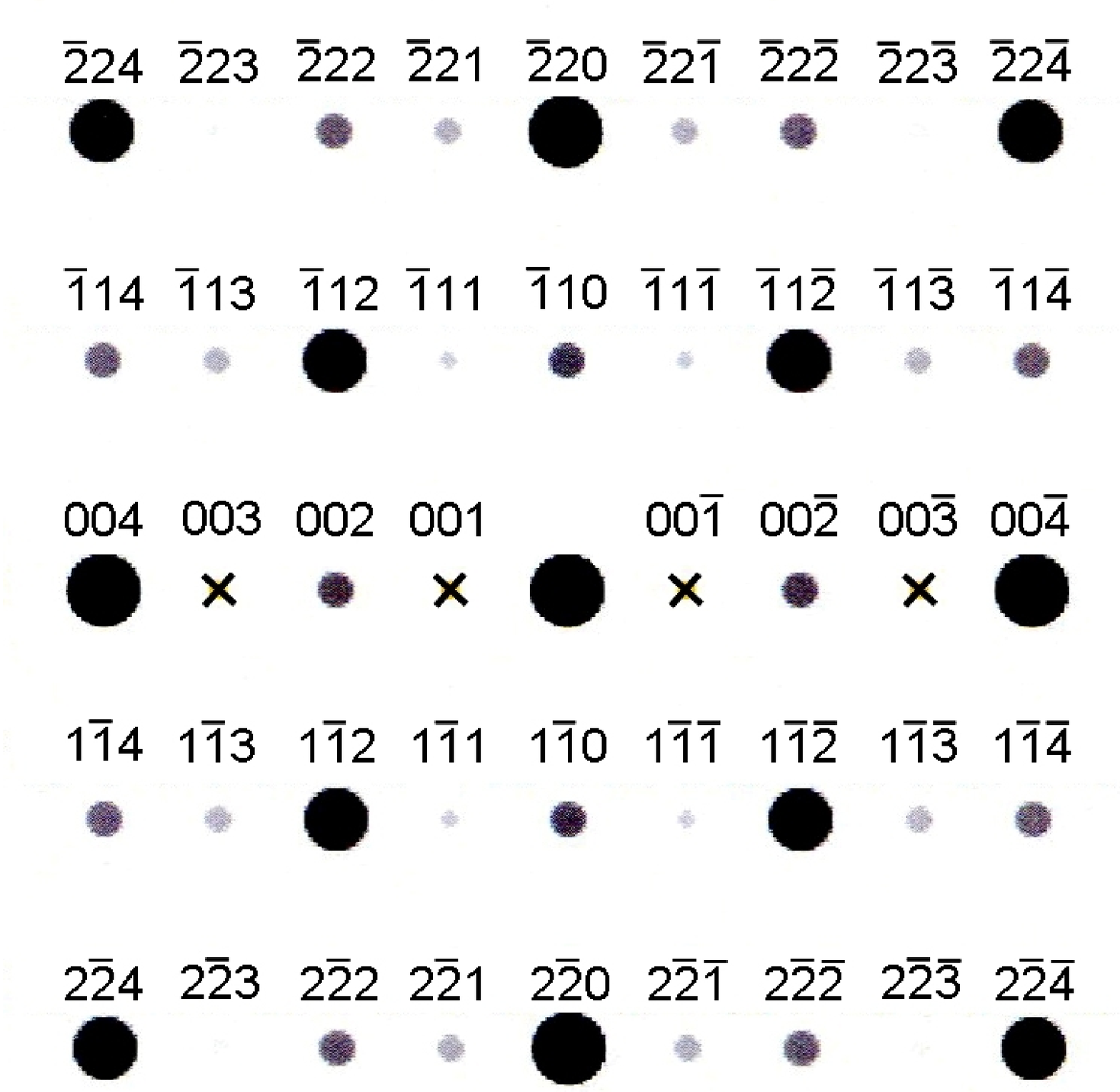}}
\caption{Simulated diffraction pattern for orthorhombic SrRuO$_3$. Zone axis $[110]_o$, i.e.~in a direction corresponding
to equivalent zone axes in Figs.~\protect{\ref{s5}} and \protect{\ref{s6}}. Kinematically forbidden but dynamically appearing
reflections are marked by crosses, among them the $[001]_o$ reflection used during reconstruction of the dark-field images
in those figures. Spot size is proportional to the intensity of the reflection. (Simulation performed by JEMS program [P. Stadelmann,
EPFL Lausanne, Switzerland]).}
\label{s7}
\end{figure}
A distinction between orthorhombic and tetragonal phases is possible for the dynamically appearing $(100)_o$, $(010)_o$ and $(001)_o$ reflections
which are present in the orthorhombic phase, but absent in the tetragonal phase.

For sample SL1, in the reconstructed dark-field image of Fig.~\ref{s3} (right) in the light of the $(010)_o$ reflection, the SRO layers are
mostly imaged with high intensity (i.e.~bright). This means that in sample SL1, the SRO layers are clearly orthorhombic, with the $c$-axis
in the plane of the layers. Opposite to this, for sample SL3, in the corresponding reconstructed dark-field image of Fig.~\ref{s4} (right)
in the light of the $(010)_o$ reflection, the SRO layers are all imaged with very low intensity (i.e.~dark). Since this could also mean
that the orientation of the SRO layers was different from the one in sample SL1, whereas still being orthorhombic, the other two possible
orientations were studied as well. As Fig.~\ref{s5} shows, the reconstructed dark-field image in the light of the $(001)_o$ reflection (right)
for the in-plane $c$-axis perpendicular to the viewing direction gives although non-zero, but still rather low intensity of the SRO layers.
Figure~\ref{s6} shows that the intensity of the SRO layers in the reconstructed dark-field image in the light of the $(001)_o$ reflection (right)
for the orthorhombic SRO unit cell with out-of-plane $c$-axis was zero. In result, the SRO layers in sample SL3 were either not orthorhombic,
or contained only a very minor proportion of the orthorhombic phase. A corresponding FFT-based analysis of the HAADF-STEM images of the same
two samples gave analogous results.

Figure~\ref{s7} shows part of a simulated diffraction pattern of the orthorhombic SRO structure along the zone axis $[110]_o$, in particular
revealing the (indicated by crosses) $[001]_o$ reflections used during reconstructed dark-field imaging in Figs.~\ref{s5} and \ref{s6}.
Different from the FFT patterns in Figs.~\ref{s3} to \ref{s6} which resulted from superpositions of the three SRO (and additionally PCMO)
orientations shown in the insets of Figs.~\ref{s4} to \ref{s6}, Fig.~\ref{s7} shows the diffraction pattern of only one single SRO orientation.
The latter corresponds (slightly rotated) to the FFT pattern and schematic inset of Fig.~\ref{s6}. Although $[001]_o$ reflections are
kinematically forbidden in the orthorhombic space group of SRO, they nevertheless appear due to dynamical diffraction conditions.

In all the Figs.~\ref{s4} to \ref{s6} the SRO layers were dark, which means that they were not orthorhombic in sample SL3. Only occasionally,
small spots of intensity could be seen in the SRO layers, which might indicate that there are very few orthorhombic domains in sample SL3.

In conclusion, the SRO layers of sample SL1 were orthorhombic, whereas those in sample SL3 were either not orthorhombic or contain only
very few orthorhombic domains.
\section{Magnetotransport properties}
\subsection{Theoretical considerations\label{theory}}
\begin{figure}[t]
\centerline{\includegraphics[width=0.7\textwidth]{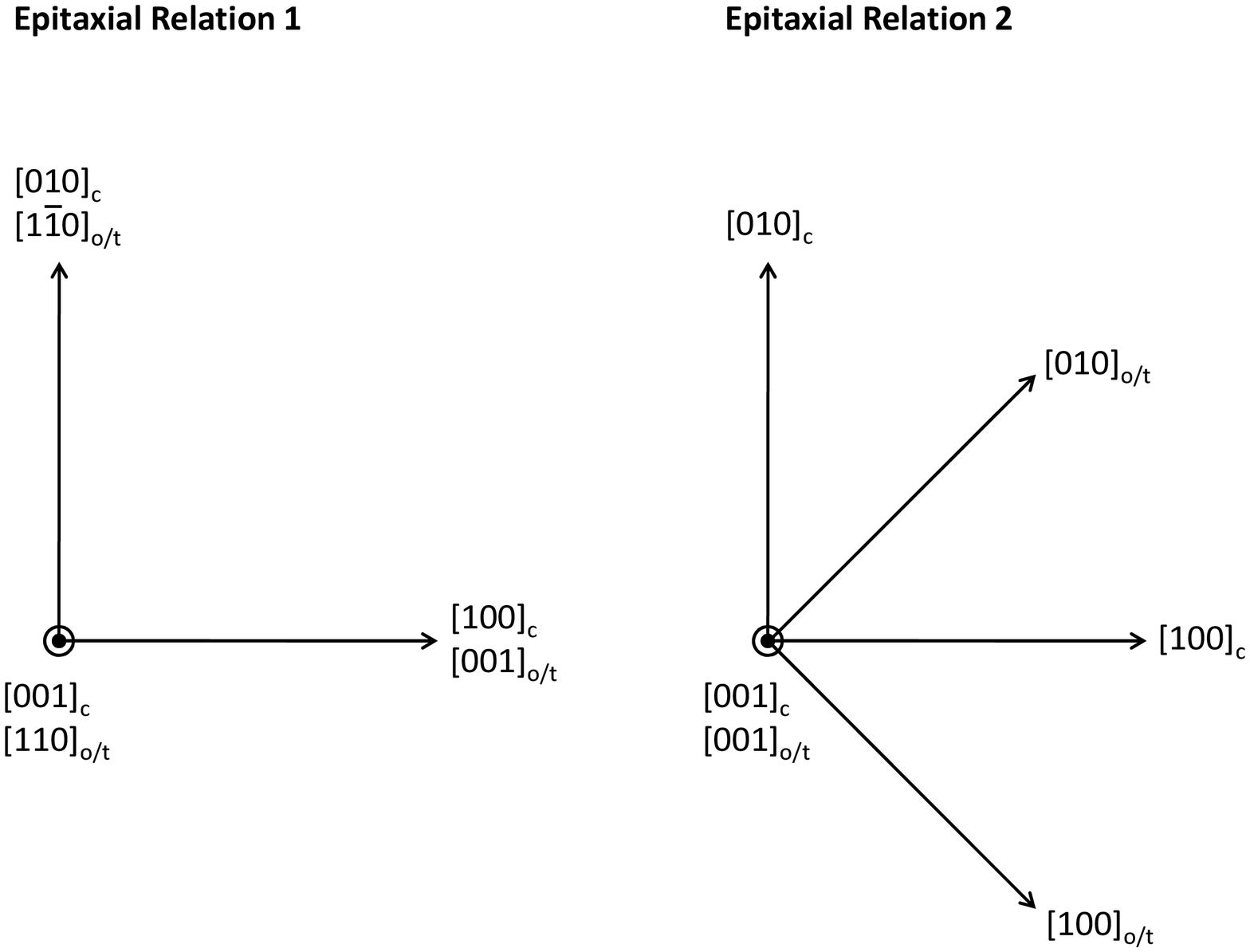}}
\caption{Sketch of the two epitaxial relations possible for the growth of orthorhombic (o) and tetragonal (t) SrRuO$_3$ films on SrTiO$_3$$(001)$ substrates.
$[001]_c$ is along the growth direction of the film.}
\label{setup}
\end{figure}

In a ferromagnet, anisotropy energy and resistivity are functions of the direction of the spontaneous magnetization.
Therefore, it is possible to conclude on the crystalline symmetry from direction-dependent measurements of the anisotropic magnetoresistance (AMR).
The relevant equations for this approach are summarized in the appendix; the derivation followed D\"oring and Simon.\cite{doering1960,doering1961}

The MR $\Delta\rho/\rho_0$ can be written as a function of a symmetric tensor of second rank
$(A_{ij})_{i,j=1,3}$ reduced by the directional unit vector $(\beta_1,\beta_2,\beta_3)$ of the current density:\cite{doering1960}
\begin{equation}
\Delta\rho/\rho_0 = \sum_{i,j=1}^3\, A_{ij}\beta_i\beta_j\, .
\label{eq1}
\end{equation}
By definition the $\beta_i$ are the direction cosines of the current density with respect to the crystallographic
basis vectors. The tensor components $A_{ij}$ are functions of the direction cosines of the magnetization,
$\vec{M} = M_S\hat{m} = M_S (\alpha_1,\alpha_2,\alpha_3)$, where $M_S$ denotes the saturation magnetization
and $\hat{m}$ the unit vector along the magnetization direction. The functional form of the matrices $(A_{ij})_{i,j=1,3}$
was obtained from crystal symmetry considerations in [\onlinecite{doering1957,doering1960,doering1961}].
Note that in case of crystal systems with a normal basis, the direction cosines obey
$\alpha_1^2+\alpha_2^2+\alpha_3^2 = \beta_1^2+\beta_2^2+\beta_3^2 =1$.

In the following indices ``c'', ``t'' and ``o'' refer to the cubic directions of SrTiO$_3$ and to the tetragonal
or orthorhombic directions of SrRuO$_3$, respectively. For SrRuO$_3$ either orthorhombic\cite{gan1999} or tetragonal\cite{vailionis2008}
symmetry was assumed. The orthorhombic cell has four times the volume of the pseudocubic cell, i.e.~in this cell the c-axis
parameter and the basal plane area are doubled compared to the pseudocubic cell. The minimal tetragonal
cell would have twice the volume of the pseudocubic cell;\cite{zakharov1999} for better comparison with the orthorhombic case,\cite{kennedy1998} however,
we chose a tetragonal cell also with four times the volume of the pseudocubic cell. We have mainly investigated two epitaxial relations for both
orthorhombic and tetragonal symmetry. In the first epitaxial relation the $[001]_{o/t}$ axis is along the substrate $[100]_c$ axis,
whereas the $[1\overline{1}0]_{o/t}$ axis is along the substrate $[010]_c$ axis, see Fig.~\ref{setup}(a). In the second epitaxial relation
the $[001]_{o/t}$ axis is along the substrate $[001]_c$ axis, wheras the $[100]_{o/t}$ and $[010]_{o/t}$ axes
are rotated with respect to the substrate $[100]_c$ and $[010]_c$ axes by $45$~degrees, see Fig.~\ref{setup}(b).
In case of single SrRuO$_3$ films grown on SrTiO$_3$$(001)$ substrates the first epitaxial relation is realized;\cite{gan1999,ziese2010c}
macroscopic alignment of the orthorhombic $[001]_o$ axis along terrace steps is achieved by growth on slightly vicinal substrates.
SrRuO$_3$ films grown on SrTiO$_3$$(001)$ in this fashion have a small monoclinic distortion with the angle between the orthorhombic
$a$- and $b$-axes deviating from a right angle by about half a degree.\cite{gan1999} Therefore in case of this epitaxial relation
also monoclinic symmetry is considered.

MR measurements were performed at constant magnetic field as a function of angle. For this the orientation of the
substrate crystal was used as a reference system and angular sweeps in the $(100)_c$, $(010)_c$ and $(001)_c$ planes were performed.
The direction of the magnetization vector with respect to the substrate crystal is specified by spherical coordinates,
$\hat{m} = (\sin\theta\cos\varphi,\sin\theta\sin\varphi,\cos\theta)$ with the angles $\theta$ and $\varphi$ defined
with respect to the $[001]_c$ and $[100]_c$ axes, respectively. Note that the magnetization angles are not necessarily
identical to the angles $\theta_F$ and $\varphi_F$ between magnetic field and the substrate axes $[001]_c$ and $[100]_c$ that were
directly measured. Accordingly the angles in out-of-plane field rotations
are specified by $\theta_F$ and in in-plane field rotations by $\varphi_F$.
The angular dependence of the anisotropic MR as determined from symmetry considerations was derived
for the two epitaxial relations and the three rotation planes. Tetragonal, orthorhombic and monoclinic crystal structures are discussed;
the relevant equations are summarized in the appendix.

\subsection{Angular magnetoresistance}
\begin{figure}[t]
\centerline{\includegraphics[width=0.95\textwidth]{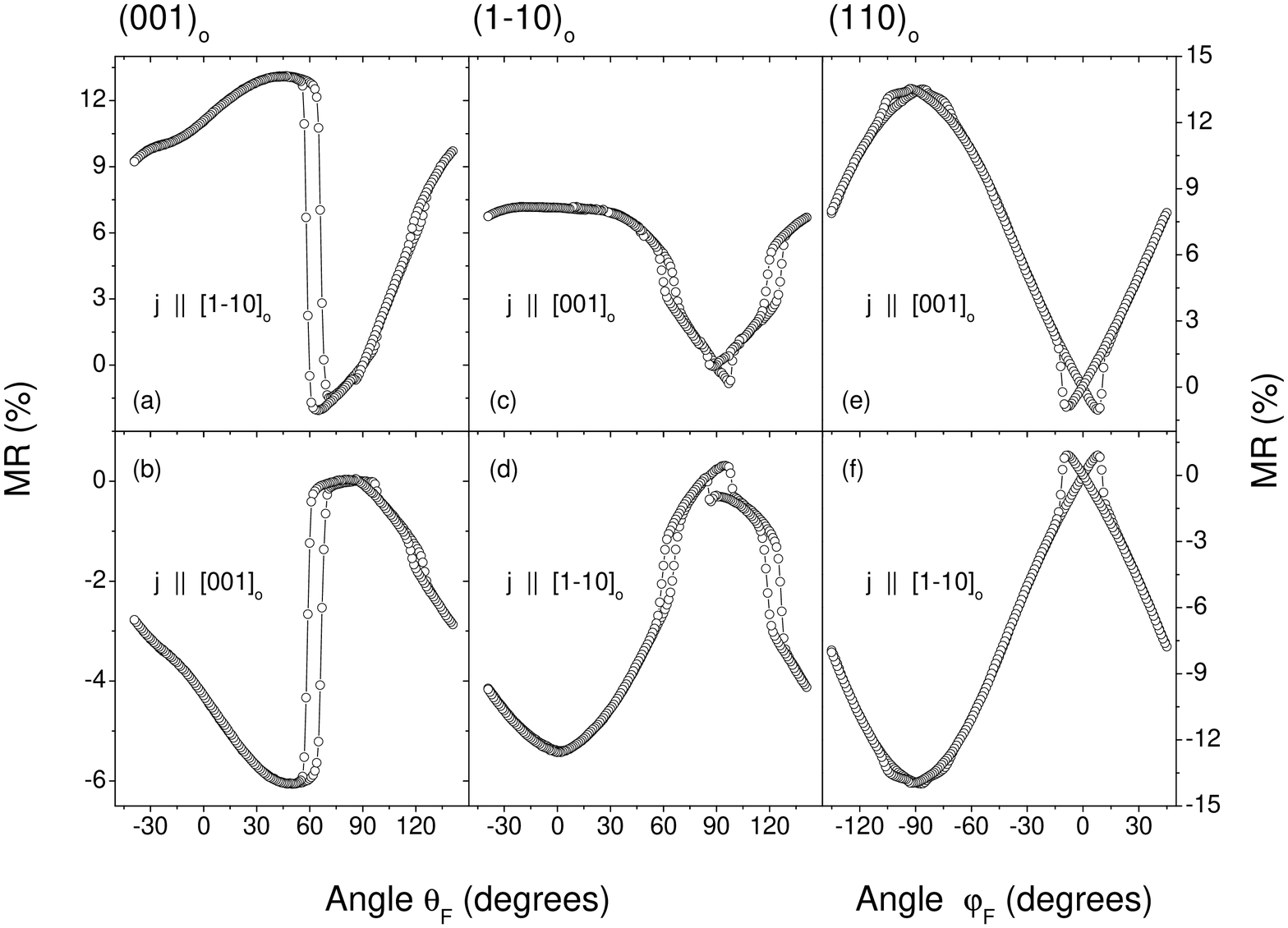}}
\caption{Sample SL1: $T = 10$~K, $\mu_0H = 8$~T. Angular dependence of the magnetoresistance for magnetic field rotation
in the $(001)_o$-, $(1\overline{1}0)_o$- and $(110)_o$-planes. The current directions are indicated in each panel.}
\label{mr1}
\end{figure}
\begin{figure}[t]
\centerline{\includegraphics[width=0.95\textwidth]{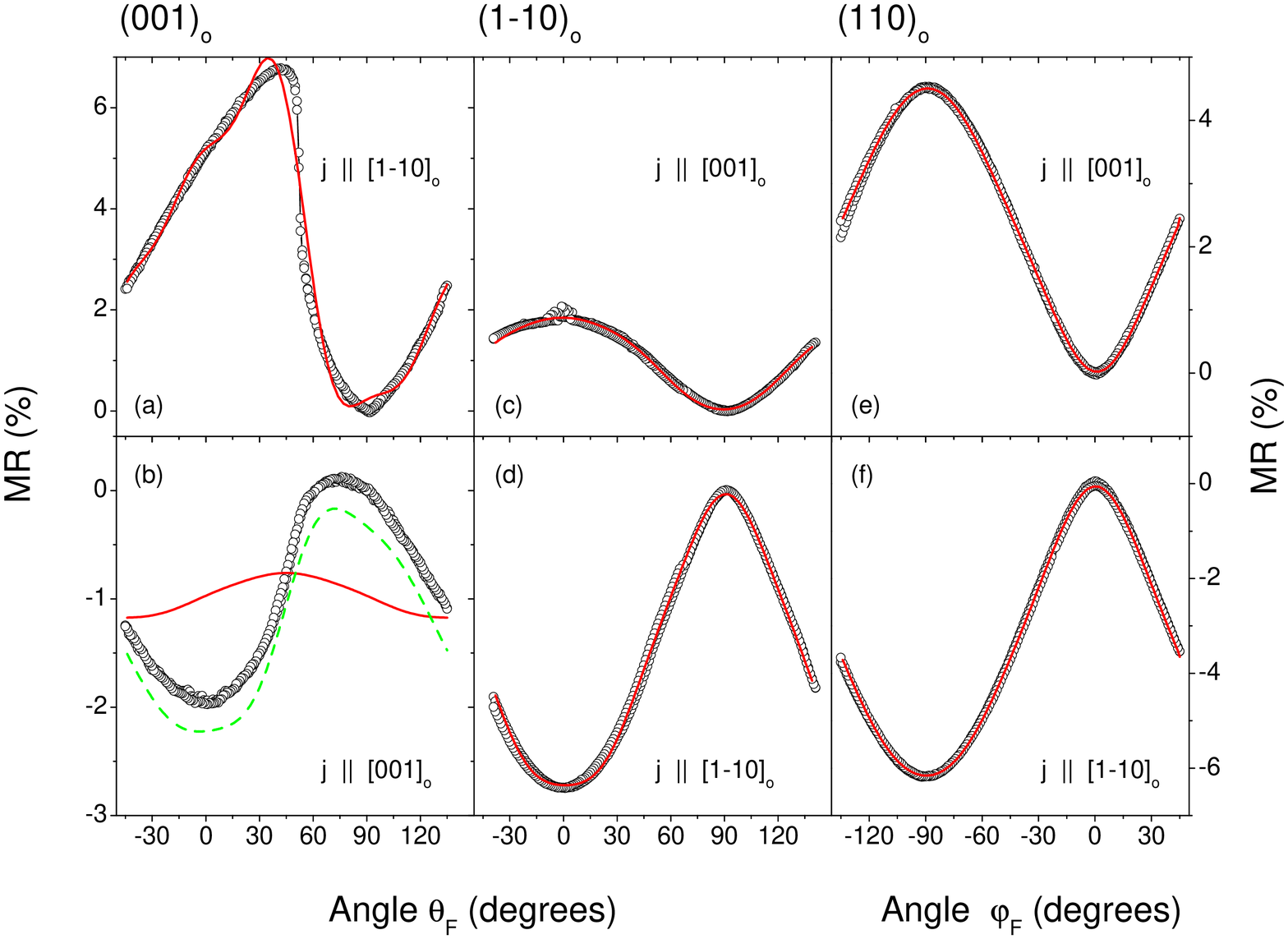}}
\caption{Sample SL1: $T = 130$~K, $\mu_0H = 8$~T. Angular dependence of the magnetoresistance for magnetic field rotation
in the $(001)_o$-, $(1\overline{1}0)_o$- and $(110)_o$-planes. The current directions are indicated in the figure.
The solid red lines are fits of Eqs.~(\protect{\ref{eq5}}-\protect{\ref{eq9}}) for orthorhombic symmetry to the data.
The dashed green curve is a fit of Eq.~(\protect{\ref{eq10}}) for monoclinic symmetry to the data; for clarity this
curve was downshifted by $-0.3$\% (absolute).}
\label{mr2}
\end{figure}
The resistivity and the angular dependent MR of the three samples shown in Table~\ref{table1} were measured.
Since PCMO single layers are insulating,\cite{ziese2010e} the resitivity and MR of the superlattices were
entirely dominated by the SRO layers. Correspondingly, the resistivity of the PCMO/SRO SLs showed a slope change
of the resistivity at the Curie temperature of the SRO layers,\cite{ziese2010e} from which the Curie
temperature of 143~K (as shown in Table~\ref{table1}) for the SRO layers was determined.
The Curie temperature of the PCMO layers of 110-115~K was determined from magnetization measurements.\cite{ziese2010d}

Here we only show angle dependent MR measurements, since these allow for the determination of the crystalline
symmetry. The measurements were performed at various temperatures between 10 and 150~K. In the following the data for samples SL1 and SL3
at 10 and 130~K are shown. The angle dependence of the MR of sample SL2 had the same form as that of sample SL3.
Figure~\ref{mr1} shows the MR of sample SL1 measured at 10~K. The MR shows hysteresis for certain angles that indicates the presence of a magnetically
hard axis close to the corresponding direction. Since SRO has a large magnetocrystalline anisotropy,\cite{ziese2010c} even at 8~T
the magnetization and magnetic-field direction do not agree at low temperatures. In case of sample SL1 the direction of two magnetically
hard axes is clearly identified. One lies in the $(001)_o$-plane at about 60 degrees from the $[110]_o$ direction,
see sharp hysteretic MR jump in Figs.~\ref{mr1}(a) and (b); the second is along the $[001]_o$ axis, see hysteresis close to $\theta_F = 0$
in Fig.~\ref{mr1}(c) and (d) and $\varphi_F = 0$ in Figs.~\ref{mr1}(e) and (f). The magnetic hard axes directions are characteristic
for {\em orthorhombic} SRO films grown on SrTiO$_3$$(001)$.\cite{kolesnik2006,ziese2010c} Further, comparing the angular MR traces of sample SL1
with the data of the 40~nm thick SrRuO$_3$ single film presented in [\onlinecite{ziese2010c}], it is immediately evident that the SRO layers
in sample SL1 have orthorhombic, actually monoclinic, symmetry. Fitting of the experimental data at 10~K is difficult, since the magnetocrystalline
anisotropy energy of the SRO layers is not accurately known and therefore the relation between magnetization angles ($\theta,\varphi)$
and magnetic field angles $(\theta_F,\varphi_F)$ is difficult to determine. However, at higher temperatures, the thermal fluctuations
are larger and the magnetocrystalline energy might be smaller and the MR curves are smooth, see the MR data of sample SL1 at 130~K in Fig.~\ref{mr2}.
Thus one might assume $\theta \simeq \theta_F$ and
$\varphi \simeq \varphi_F$. Eqs.~(\ref{eq5}-\ref{eq9}) derived for orthorhombic symmetry accurately fit the data
in Figs.~\ref{mr2}(a) and (c)-(f), but not (b), see solid red lines; for the fitting the expressions Eqs.~(\ref{eq5}-\ref{eq9}) were truncated at eighth order.
This is in agreement with the results for a single SRO film in [\onlinecite{ziese2010c}]. The MR curve in Fig.~\ref{mr2}(b) cannot be described
by Eq.~(\ref{eq6}) even if higher order terms were taken into account, since the experimental data contain a large $\cos(2\theta)$ term
absent in Eq.~(\ref{eq6}). The corresponding expression for monoclinic symmetry, Eq.~(\ref{eq10}), contains this term and fits
the data in Fig.~\ref{mr2}(b) well, see dashed green line. This result is again in agreement with the SRO single film results\cite{ziese2010c}
and is consistent with the fact that a monoclinic distortion between $a$- and $b$-axes was observed in SRO films.\cite{gan1999,vailionis2007}
Since the MR data cannot be understood within the other epitaxial relations and crystalline symmetries, from the MR analysis
we firmly conclude that the crystalline symmetry of the SRO layers of sample SL1 is orthorhombic (monoclinic). This is in full
agreement with the HRTEM results discussed in section~\ref{tem}.

\begin{figure}[t]
\centerline{\includegraphics[width=0.95\textwidth]{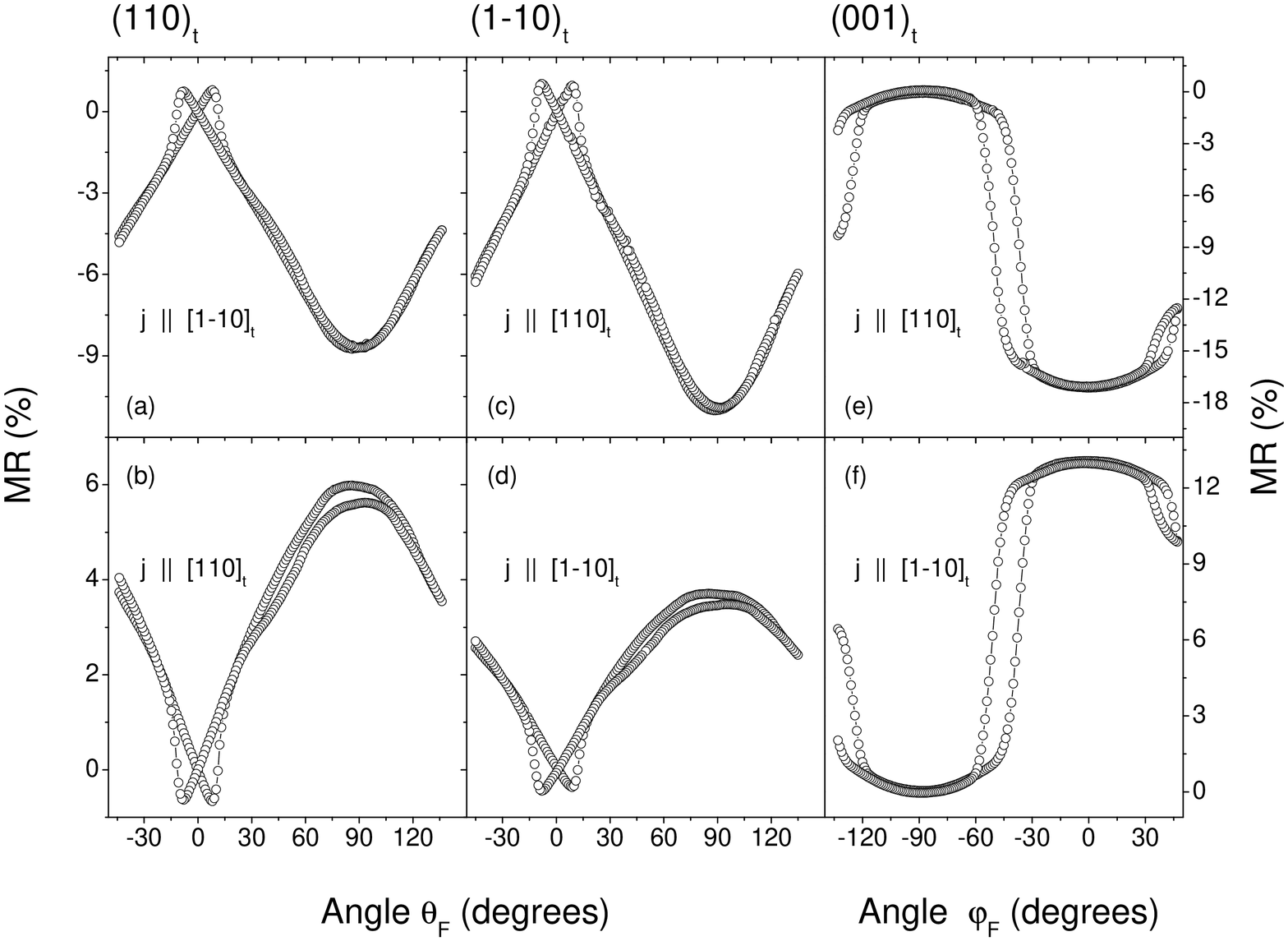}}
\caption{Sample SL3: $T = 10$~K, $\mu_0H = 8$~T. Angular dependence of the magnetoresistance for magnetic field rotation
in the $(110)_t$-, $(1\overline{1}0)_t$- and $(001)_t$-planes. The current directions are indicated in each panel.}
\label{mr3}
\end{figure}
\begin{figure}[t]
\centerline{\includegraphics[width=0.95\textwidth]{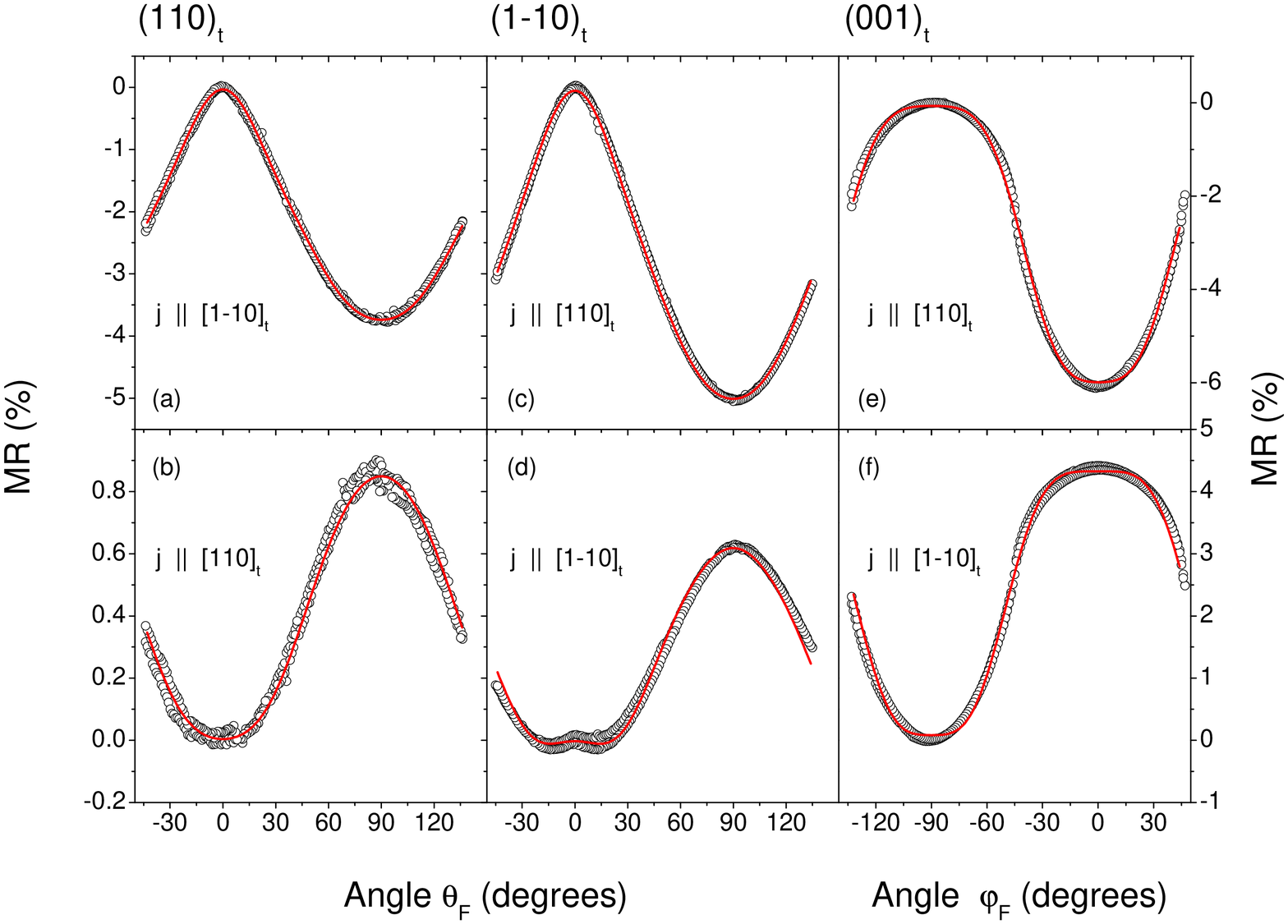}}
\caption{Sample SL3: $T = 130$~K, $\mu_0H = 8$~T. Angular dependence of the magnetoresistance for magnetic field rotation
in the $(110)_t$-, $(1\overline{1}0)_t$- and $(001)_t$-planes. The current directions are indicated in the figure.
The solid red lines are fits of Eqs.~(\protect{\ref{eq11}}-\protect{\ref{eq13}}) for tetragonal symmetry to the data.}
\label{mr4}
\end{figure}
Figure~\ref{mr3} shows the angular MR of sample SL3 measured at 10~K. Comparison with Fig.~\ref{mr1} shows that
the form of the angular dependence is significantly different from the orthorhombic case with the $c$-axis in-plane.
In case of sample SL3 hysteresis occurs close to the surface normal, $\theta_F = 0$~degrees, see Figs.~\ref{mr3}(a)-(d)
and in case of in-plane rotations near $\varphi_F = -45$~degrees, see Figs.~\ref{mr3}(e)-(f). Accordingly, compared to sample
SL1 the SRO layers in sample SL3 have another orientation, another crystalline symmetry or both. Since the form
of the MR curves in Figs.~\ref{mr3}(a) and (c) as well as (b) and (d) are very similar, it appears probable that the $c$-axis
of the either tetragonal or orthorhombic structure is along the SL normal, i.e.~that the second epitaxial relation
shown in Fig.~\ref{setup} is realized for sample SL3. The out-of-plane rotations shown in Fig.~\ref{mr3}(a)-(d), however,
do not allow for a discrimination of orthorhombic and tetragonal symmetry, since these rotations occur
in the $[110]_{o/t}$ and $[-110]_{o/t}$ planes that have equivalent symmetry in the two crystallographic structures.

Since the magnetocrystalline anisotropy energy of the SRO layers in sample SL3 is very large, fits of Eqs.~(\ref{eq11}-\ref{eq13})
were made to the MR data of sample SL3 at 130~K. These fits are shown by the red lines in Fig.~\ref{mr4}; as above
the expressions were truncated at eighth order. The fitting of the curves to the data is fully convincing. However, since the functional
form of the MR in Eqs.~(\ref{eq11}-\ref{eq13}) is the same for orthorhombic and tetragonal symmetry, this agreement
does not yet discriminate between the two crystalline structures. A discrimination is, however, possible by an analysis of
the expansion coefficients in the expressions for the in-plane rotation. In case of tetragonal symmetry, the coefficients $c^{t1}_{2n}$
and $c^{t2}_{2n}$ for the two current-density directions have a definite relationship: $c^{t2}_{4n} = c^{t1}_{4n},\, c^{t2}_{4n-2} = -c^{t1}_{4n-2},\, n = 1,2,3...$,
whereas the corresponding coefficients for the orthorhombic symmetry are independent of each other.
The coefficients obtained at 130~K are shown for samples SL2 and SL3 as well as -- for comparison -- for sample SL1 in Table~\ref{table2}.
In case of the first two samples the coefficients indeed show the alternating $+/-$ pattern as predicted for tetragonal symmetry,
see Eqs.~(\ref{eq12},\ref{eq13}), whereas the corresponding coefficients obtained for sample SL1 do not show this pattern, see rightmost
two columns of Table~\ref{table2}. Therefore we conclude from the angular dependent MR measurements that the SRO layers
in samples SL2 and SL3 have tetragonal symmetry with the $c$-axis along the SL normal. This conclusion is consistent with the
conclusion of the HRTEM studies that the SRO layers in sample SL3 are not orthorhombic, see section~\ref{tem}.
We cannot firmly exclude, however, the existence of tetragonal domains with an in-plane $c$-axis.

\begin{table}
\caption{Coefficients of Eqs.~(\protect{\ref{eq12}},\protect{\ref{eq13}}) for samples SL3 and SL2 and of
Eqs.~(\protect{\ref{eq8}},\protect{\ref{eq9}}) for sample SL1 at 130~K.}
\label{table2}
\begin{tabular}{l|l|l|l|l|l|l}
\hline
 & SL2 & & SL3 & & SL1 & \\
$2n$ & $c^{t1}_{2n}$ & $c^{t2}_{2n}$ & $c^{t1}_{2n}$ & $c^{t2}_{2n}$ & $c^{o1}_{2n}$ & $c^{o2}_{2n}$\\
\hline
$0$ & $-229\times10^{-4}$  & $+341\times10^{-4}$  & $-280\times10^{-4}$  & $+239\times10^{-4}$  & $+237.8\times10^{-4}$ & $-341.0\times10^{-4}$\\
$2$ & $-271\times10^{-4}$  & $+349\times10^{-4}$  & $-328\times10^{-4}$  & $+238\times10^{-4}$  & $-216.0\times10^{-4}$ & $+296.9\times10^{-4}$\\
$4$ & $-25.3\times10^{-4}$ & $-26.3\times10^{-4}$ & $-28.7\times10^{-4}$ & $-24.3\times10^{-4}$ & $-9.5\times10^{-4}$   & $+27.4\times10^{-4}$\\
$6$ & $+30.0\times10^{-4}$ & $-40.9\times10^{-4}$ & $+31.9\times10^{-4}$ & $-25.8\times10^{-4}$ & $-8.3\times10^{-4}$   & $+8.5\times10^{-4}$\\
$8$ & $+7.3\times10^{-4}$  & $+6.2\times10^{-4}$  & $+6.0\times10^{-4}$  & $+5.4\times10^{-4}$  & $-2.0\times10^{-4}$   & $+3.0\times10^{-4}$\\
\hline
\end{tabular}

\end{table}

\section{Discussion and conclusions\label{summary}}
In this work we have shown by a combination of two techniques, namely high-resolution transmission electron microscopy and angular
dependent magnetoresistance measurements, that the SrRuO$_3$ layers in a Pr$_{0.7}$Ca$_{0.3}$MnO$_3$/SrRuO$_3$ superlattice
undergo a phase transition from orthorhombic to tetragonal structure, when the thickness of the PCMO layers is increased
from 1.5 to 4~nm. The orthorhombic $c$-axis of the SRO layers was found to lie in-plane, whereas the tetragonal $c$-axis of the
SRO layers seemed to be oriented along the SL normal; for the tetragonal orientation, however, the existence of crystallographic
domains with in-plane $c$-axis cannot be fully excluded. The results impressively demonstrate that this structural phase transition
has a large impact on the magnetotransport properties. This is surprising, since the actual atomic displacements between the phases are rather small.

What drives this phase transition? An obvious candidate in case of thin films and superlattices is the strain. Indeed, the transition
temperature for the orthorhombic(O)-to-tetragonal(T) transition was found to be substantially lowered in compressively strained SRO films
grown on SrTiO$_3$$(001)$.\cite{vailionis2007,choi2010} Since PCMO has an even smaller lattice constant than STO, a further lowering
of the OT-transition temperature might be expected; furthermore, the strain exerted by the PCMO layers is not biaxial, but
anisotropic, which might also modify the strain effect. Since we observed the tetragonal structure of the SRO layers down to 10~K,
this would mean that the strain effect had lowered the OT-transition temperature basically to zero. Although this scenario is not excluded,
it might appear unlikely. In an alternative scenario the OT-transition might be influenced by the electronic or magnetic coupling between
the PCMO and SRO layers. Here it would be interesting for future research to look for structural anomalies in these superlattices near the
magnetic transition at 110~K and the charge ordering transition near 240~K.\cite{tomioka1996}

\acknowledgments This work was supported by the German Science Foundation (DFG) within the Collaborative Research Center SFB 762
``Functionality of Oxide Interfaces''. We thank Dr.~J.~Henk for a careful reading of the manuscript.

\section*{Appendix}

The unit vector of the magnetization in the system of the SrTiO$_3$ substrate crystal is written as
$\hat{m} = (\sin\theta\cos\varphi,\sin\theta\sin\varphi,\cos\theta)$ such that $\theta$ is the angle between the magnetization and $[001]_c$
and $\varphi$ is the angle between the magnetization and $[100]_c$. The direction cosines of the magnetization $(\alpha_1,\alpha_2,\alpha_3)$
and the current density $(\beta_1,\beta_2,\beta_3)$ are defined with respect to the crystallographic axes of the SRO film.

The formulas below just indicate the structure of the solutions. Unless indicated otherwise the coefficients in the equations
-- although throughout denoted by $c_{2n}$ and $s_{2n}$ -- are different
for the various rotation planes and current directions. Miller indices $(hkl)$ specify the rotation plane, the direction vector $[uvw]$
specifies the corresponding current density direction.

\subsection{Epitaxial Relation 1, tetragonal symmetry (D$_{4h}$)}

\subsubsection{
$(100)_c$/$(001)_t$, $[1\overline{1}0]_t$.
$(010)_c$/$(1\overline{1}0)_t$, $[001]_t$.
$(010)_c$/$(1\overline{1}0)_t$, $[1\overline{1}0]_t$.}

\begin{equation}
\Delta\rho/\rho_0 = c_0 + \sum_{n=1}^\infty\, c_{2n}\cos(2n\theta)\, .
\label{eq2}
\end{equation}

\subsubsection{$(100)_c$/$(001)_t$, $[001]_t$.}

\begin{equation}
\Delta\rho/\rho_0 = c_0 + \sum_{n=1}^\infty\, c_{4n}\cos(4n\theta)\, .
\label{eq3}
\end{equation}

\subsubsection{
$(001)_c$/$(110)_t$, $[001]_t$.
$(001)_c$/$(110)_t$, $[1\overline{1}0]_t$.}

\begin{equation}
\Delta\rho/\rho_0 = c_0 + \sum_{n=1}^\infty\, c_{2n}\cos(2n\varphi)\, .
\label{eq4}
\end{equation}

\subsection{Epitaxial Relation 1, orthorhombic symmetry (D$_{2h}$)}

\subsubsection{$(100)_c$/$(001)_o$, $[1\overline{1}0]_o$.}

\begin{equation}
\Delta\rho/\rho_0 = c_0 + \sum_{n=1}^\infty\, s_{2n}\sin(2n\theta) + \sum_{n=1}^\infty\, c_{2n}\cos(2n\theta)\, .
\label{eq5}
\end{equation}

\subsubsection{$(100)_c$/$(001)_o$, $[001]_o$.}

\begin{equation}
\Delta\rho/\rho_0 = c_0 + \sum_{n=1}^\infty\, s_{4n-2}\sin\left((4n-2)\theta\right)+\sum_{n=1}^\infty\, c_{4n}\cos(4n\theta)\, .
\label{eq6}
\end{equation}

\subsubsection{
$(010)_c$/$(1\overline{1}0)_o$, $[001]_o$.
$(010)_c$/$(1\overline{1}0)_o$, $[1\overline{1}0]_o$.}

\begin{equation}
\Delta\rho/\rho_0 = c_0 + \sum_{n=1}^\infty\, c_{2n}\cos(2n\theta)\, .
\label{eq7}
\end{equation}

\subsubsection{
$(001)_c$/$(110)_o$, $[001]_o$.}

\begin{equation}
\Delta\rho/\rho_0 = c^{o1}_0 + \sum_{n=1}^\infty\, c^{o1}_{2n}\cos(2n\varphi)\, .
\label{eq8}
\end{equation}

\subsubsection{
$(001)_c$/$(110)_o$, $[1\overline{1}0]_o$.}

\begin{equation}
\Delta\rho/\rho_0 = c^{o2}_0 + \sum_{n=1}^\infty\, c^{o2}_{2n}\cos(2n\varphi)\, .
\label{eq9}
\end{equation}

\subsection{Epitaxial Relation 1, monoclinic symmetry (C$_{2h}$)}

Compared to the orthorhombic symmetry there is only one modification:

\subsubsection{$(100)_c$/$(001)_m$, $[001]_m$.}

\begin{equation}
\Delta\rho/\rho_0 = c_0 + \sum_{n=1}^\infty\, s_{2n}\sin(2n\theta) + \sum_{n=1}^\infty\, c_{2n}\cos(2n\theta)\, .
\label{eq10}
\end{equation}

\subsection{Epitaxial Relation 2, tetragonal symmetry (D$_{4h}$)}

\subsubsection{
$(100)_c$/$(110)_t$, $[\overline{1}10]_t$.
$(100)_c$/$(110)_t$, $[110]_t$.
$(010)_c$/$(\overline{1}10)_t$, $[110]_t$.
$(010)_c$/$(\overline{1}10)_t$, $[\overline{1}10]_t$.}

\begin{equation}
\Delta\rho/\rho_0 = c_0 + \sum_{n=1}^\infty\, c_{2n}\cos(2n\theta)\, .
\label{eq11}
\end{equation}

Note that the coefficients in the expression for the first and third as well as the second and fourth configuration
are the same.

\subsubsection{$(001)_c$/$(001)_t$, $[110]_t$.}

\begin{equation}
\Delta\rho/\rho_0 = c^{t1}_0 + \sum_{n=1}^\infty\, c^{t1}_{2n}\cos(2n\varphi)\, .
\label{eq12}
\end{equation}

\subsubsection{$(001)_c$/$(001)_t$, $[\overline{1}10]_t$.}

\begin{equation}
\Delta\rho/\rho_0 = c^{t2}_0 + \sum_{n=1}^\infty\, c^{t2}_{2n}\cos(2n\varphi)\, .
\label{eq13}
\end{equation}

Note that $c^{t2}_{4n} = c^{t1}_{4n},\, c^{t2}_{4n-2} = -c^{t1}_{4n-2},\, n = 1,2,3...$.

\subsection{Epitaxial Relation 2, orthorhombic symmetry (D$_{2h}$)}

For orthorhombic symmetry the equations for the out-of-plane rotation configurations have the same
structure as the corresponding ones for tetragonal symmetry specified in the preceding section. This comes from the fact that the choice of rotation
planes to be the $(100)_c$ and $(010)_c$ planes does not allow for a discrimination of the orthorhombic $a$- and $b$-axes
which are under 45 degrees with respect to the rotation planes. The symmetry of the in-plane rotation, however, leads
to different expansions.

\subsubsection{
$(001)_c$/$(001)_o$, $[110]_o$.
$(001)_c$/$(001)_o$, $[\overline{1}10]_o$.}

\begin{equation}
\Delta\rho/\rho_0 = c_0 + \sum_{n=1}^\infty\, s_{2n}\sin(2n\varphi) + \sum_{n=1}^\infty\, c_{2n}\cos(2n\varphi)\, .
\label{eq14}
\end{equation}

\clearpage

%

\end{document}